\documentclass[final,3p,times]{elsarticle}

\usepackage{amssymb}
\usepackage{color}
\usepackage{verbatim}
\usepackage{amsmath}

\usepackage[hidelinks,colorlinks=true,allcolors=blue,urlcolor=blue]{hyperref}

\usepackage{lineno}

\newcommand{\car}{\mathrm{nat}}
\newcommand{\D}{\mathrm{d}}
\newcommand{\vrt}{ {\color{white}\large |} }

\journal{Physics Letters B}

\begin{document}

\begin{frontmatter}



\title{Measurement of the  $^\car$C(\textit{n,p}) and $^\car$C(\textit{n,d}) reactions from n\_TOF at CERN}

\tnotetext[n_TOF]{The n\_TOF Collaboration (www.cern.ch/ntof)}


\author[12]{ P.~\v{Z}ugec\corref{corr}}
\ead{pzugec@phy.hr}
\cortext[corr]{Corresponding author}
\author[10]{ N.~Colonna } %
\author[23]{ D.~Rochman } %
\author[1,10]{ M.~Barbagallo } %
\author[5]{ J.~Andrzejewski } %
\author[5]{ J.~Perkowski } %
\author[34]{ A.~Ventura } %
\author[12]{ D.~Bosnar } %
\author[5]{ A.~Gawlik-Ramiega } %
\author[18,1]{ M.~Sabat\'{e}-Gilarte } %
\author[1,8,9]{ M.~Bacak } %
\author[1]{ F.~Mingrone } %
\author[1,11]{ E.~Chiaveri } %
\author[1]{ O.~Aberle } %
\author[2]{ V.~Alcayne } %
\author[3,4]{ S.~Amaducci } %
\author[6]{ L.~Audouin } %
\author[7]{ V.~Babiano-Suarez } %
\author[11]{ S.~Bennett } %
\author[9]{ E.~Berthoumieux } %
\author[11]{ J.~Billowes } %
\author[13]{ A.~Brown } %
\author[14,15]{ M.~Busso } %
\author[16]{ M.~Caama\~{n}o } %
\author[7]{ L.~~Caballero-Ontanaya } %
\author[17]{ F.~Calvi\~{n}o } %
\author[1]{ M.~Calviani } %
\author[2]{ D.~Cano-Ott } %
\author[17]{ A.~Casanovas } %
\author[1]{ F.~Cerutti } %
\author[17]{ G.~Cort\'{e}s } %
\author[18]{ M.~A.~Cort\'{e}s-Giraldo } %
\author[3]{ L.~Cosentino } %
\author[14,19]{ S.~Cristallo } %
\author[10,20]{ L.~A.~Damone } %
\author[11]{ P.~J.~Davies } %
\author[21,1]{ M.~Diakaki } %
\author[22]{ M.~Dietz } %
\author[7]{ C.~Domingo-Pardo } %
\author[23]{ R.~Dressler } %
\author[24]{ Q.~Ducasse } %
\author[9]{ E.~Dupont } %
\author[16]{ I.~Dur\'{a}n } %
\author[25]{ Z.~Eleme } %
\author[16]{ B.~Fern\'{a}ndez-Dom\'{\i}nguez } %
\author[1]{ A.~Ferrari } %
\author[3]{ P.~Finocchiaro } %
\author[26]{ V.~Furman } %
\author[27]{ K.~G\"{o}bel } %
\author[22]{ R.~Garg } %
\author[1]{ S.~Gilardoni } %
\author[28]{ I.~F.~Gon\c{c}alves } %
\author[2]{ E.~Gonz\'{a}lez-Romero } %
\author[18]{ C.~Guerrero } %
\author[9]{ F.~Gunsing } %
\author[29]{ H.~Harada } %
\author[23]{ S.~Heinitz } %
\author[30]{ J.~Heyse } %
\author[13]{ D.~G.~Jenkins } %
\author[31]{ A.~Junghans } %
\author[32]{ F.~K\"{a}ppeler\fnref{passed} } %
   \fntext[passed]{Passed away in 2021.}
\author[1]{ Y.~Kadi } %
\author[29]{ A.~Kimura } %
\author[33]{ I.~Knapov\'{a} } %
\author[21]{ M.~Kokkoris } %
\author[26]{ Y.~Kopatch } %
\author[33]{ M.~Krti\v{c}ka } %
\author[27]{ D.~Kurtulgil } %
\author[7]{ I.~Ladarescu } %
\author[22]{ C.~Lederer-Woods } %
\author[8]{ H.~Leeb } %
\author[18]{ J.~Lerendegui-Marco } %
\author[22]{ S.~J.~Lonsdale } %
\author[1]{ D.~Macina } %
\author[34,35]{ A.~Manna } %
\author[2]{ T.~Mart\'{\i}nez } %
\author[36]{ M.~Mart\'{\i}nez-Ca\~{n}ada } %
\author[1]{ A.~Masi } %
\author[34,35]{ C.~Massimi } %
\author[37]{ P.~Mastinu } %
\author[1]{ M.~Mastromarco } %
\author[23]{ E.~A.~Maugeri } %
\author[10,38]{ A.~Mazzone } %
\author[2]{ E.~Mendoza } %
\author[39]{ A.~Mengoni } %
\author[21,1]{ V.~Michalopoulou } %
\author[40]{ P.~M.~Milazzo } %
\author[9]{ J.~Moreno-Soto } %
\author[41,4]{ A.~Musumarra } %
\author[42]{ A.~Negret } %
\author[24]{ R.~Nolte } %
\author[36]{ F.~Og\'{a}llar } %
\author[42]{ A.~Oprea } %
\author[25]{ N.~Patronis } %
\author[43]{ A.~Pavlik } %
\author[42]{ C.~Petrone } %
\author[10,14,19]{ L.~Piersanti } %
\author[24]{ E.~Pirovano } %
\author[36]{ I.~Porras } %
\author[36]{ J.~Praena } %
\author[18]{ J.~M.~Quesada } %
\author[6]{ D.~Ramos } %
\author[44,45]{ T.~Rauscher } %
\author[27]{ R.~Reifarth } %
\author[1]{ C.~Rubbia } %
\author[46]{ A.~Saxena } %
\author[30]{ P.~Schillebeeckx } %
\author[23]{ D.~Schumann } %
\author[11]{ A.~Sekhar } %
\author[11]{ A.~G.~Smith } %
\author[11]{ N.~V.~Sosnin } %
\author[23]{ P.~Sprung } %
\author[21]{ A.~Stamatopoulos } %
\author[10]{ G.~Tagliente } %
\author[7]{ J.~L.~Tain } %
\author[17]{ A.~Tarife\~{n}o-Saldivia } %
\author[1,21,6]{ L.~Tassan-Got } %
\author[27]{ Th.~Thomas } %
\author[36]{ P.~Torres-S\'{a}nchez } %
\author[1]{ A.~Tsinganis } %
\author[23]{ J.~Ulrich } %
\author[31,1]{ S.~Urlass } %
\author[33]{ S.~Valenta } %
\author[34,35]{ G.~Vannini } %
\author[10]{ V.~Variale } %
\author[28]{ P.~Vaz } %
\author[14]{ D.~Vescovi } %
\author[1]{ V.~Vlachoudis } %
\author[21]{ R.~Vlastou } %
\author[47]{ A.~Wallner } %
\author[22]{ P.~J.~Woods } %
\author[11]{ T.~Wright } %

\author[]{The n\_TOF Collaboration}


\address[12]{Department of Physics, Faculty of Science, University of Zagreb, Zagreb, Croatia} %
\address[10]{Istituto Nazionale di Fisica Nucleare, Sezione di Bari, Italy} %
\address[23]{Paul Scherrer Institut (PSI), Villigen, Switzerland} %
\address[1]{European Organization for Nuclear Research (CERN), Switzerland} %
\address[5]{University of Lodz, Poland} %
\address[34]{Istituto Nazionale di Fisica Nucleare, Sezione di Bologna, Italy} %
\address[18]{Universidad de Sevilla, Spain} %
\address[8]{TU Wien, Atominstitut, Stadionallee 2, 1020 Wien, Austria} %
\address[9]{CEA Irfu, Universit\'{e} Paris-Saclay, F-91191 Gif-sur-Yvette, France} %
\address[11]{University of Manchester, United Kingdom} %
\address[2]{Centro de Investigaciones Energ\'{e}ticas Medioambientales y Tecnol\'{o}gicas (CIEMAT), Spain} %
\address[3]{INFN Laboratori Nazionali del Sud, Catania, Italy} %
\address[4]{Department of Physics and Astronomy, University of Catania, Italy} %
\address[6]{Institut de Physique Nucl\'{e}aire, CNRS-IN2P3, Univ. Paris-Sud, Universit\'{e} Paris-Saclay, F-91406 Orsay Cedex, France} %
\address[7]{Instituto de F\'{\i}sica Corpuscular, CSIC - Universidad de Valencia, Spain} %
\address[13]{University of York, United Kingdom} %
\address[14]{Istituto Nazionale di Fisica Nucleare, Sezione di Perugia, Italy} %
\address[15]{Dipartimento di Fisica e Geologia, Universit\`{a} di Perugia, Italy} %
\address[16]{University of Santiago de Compostela, Spain} %
\address[17]{Universitat Polit\`{e}cnica de Catalunya, Spain} %
\address[19]{Istituto Nazionale di Astrofisica - Osservatorio Astronomico d'Abruzzo, Italy} %
\address[20]{Dipartimento Interateneo di Fisica, Universit\`{a} degli Studi di Bari, Italy} %
\address[21]{National Technical University of Athens, Greece} %
\address[22]{School of Physics and Astronomy, University of Edinburgh, United Kingdom} %
\address[24]{Physikalisch-Technische Bundesanstalt (PTB), Bundesallee 100, 38116 Braunschweig, Germany} %
\address[25]{University of Ioannina, Greece} %
\address[26]{Affiliated with an institute covered by a cooperation agreement with CERN} %
\address[27]{Goethe University Frankfurt, Germany} %
\address[28]{Instituto Superior T\'{e}cnico, Lisbon, Portugal} %
\address[29]{Japan Atomic Energy Agency (JAEA), Tokai-Mura, Japan} %
\address[30]{European Commission, Joint Research Centre (JRC), Geel, Belgium} %
\address[31]{Helmholtz-Zentrum Dresden-Rossendorf, Germany} %
\address[32]{Karlsruhe Institute of Technology, Campus North, IKP, 76021 Karlsruhe, Germany} %
\address[33]{Charles University, Prague, Czech Republic} %
\address[35]{Dipartimento di Fisica e Astronomia, Universit\`{a} di Bologna, Italy} %
\address[36]{University of Granada, Spain} %
\address[37]{INFN Laboratori Nazionali di Legnaro, Italy} %
\address[38]{Consiglio Nazionale delle Ricerche, Bari, Italy} %
\address[39]{Agenzia nazionale per le nuove tecnologie, l'energia e lo sviluppo economico sostenibile (ENEA), Italy} %
\address[40]{Istituto Nazionale di Fisica Nucleare, Sezione di Trieste, Italy} %
\address[41]{Istituto Nazionale di Fisica Nucleare, Sezione di Catania, Italy} %
\address[42]{Horia Hulubei National Institute of Physics and Nuclear Engineering, Romania} %
\address[43]{University of Vienna, Faculty of Physics, Vienna, Austria} %
\address[44]{Department of Physics, University of Basel, Switzerland} %
\address[45]{Centre for Astrophysics Research, University of Hertfordshire, United Kingdom} %
\address[46]{Bhabha Atomic Research Centre (BARC), India} %
\address[47]{Australian National University, Canberra, Australia} %



\begin{abstract}

The energy dependence of the cross section of the (\textit{n,p}) and (\textit{n,d}) reactions on $^\car$C has been studied for the first time at the n\_TOF facility at CERN, from the particle detection threshold up to 25~MeV. The measurement was performed with two telescopes made of position-sensitive silicon $\Delta E$-$E$ detectors, covering the angular range from 20$^\circ$ to 140$^\circ$. A detector efficiency has been determined by means of Monte Carlo simulations of the experimental setup. Various assumptions on the angular distributions and branching ratios of the excited levels of the residual $^{11}$B, $^{12}$B, $^{13}$B nuclei were considered. In particular, theoretical calculations based on the TALYS-2.0 code were used and the systematic uncertainties in the analysis results were determined from the variations in these distributions. The n\_TOF data on the (\textit{n,p}) and (\textit{n,d}) reaction on carbon are characterized by a higher accuracy and wider energy range than currently available in literature. A comparison with current evaluations from different libraries reveals a rather significant disagreement with the n\_TOF results, while a remarkable agreement is observed with the prediction of TALYS-2.0 for this light element.

\end{abstract}




\end{frontmatter}



\section{Introduction}
\label{introduction}

The $^\car$C(\textit{n},cp) reactions with charged particles (cp) in the exit channel are among the most important ones for a variety of fields. Together with oxygen, nitrogen and other light elements, carbon is abundantly present in human body, and knowledge of the cross section of neutron-induced reactions is important for estimates of dose to tissues in hadrontherapy, in particular with protons and light ion beams. Although the importance of neutron induced reactions on carbon in nuclear medicine has long been recognized~\cite{chadwick}, data and evaluations on (\textit{n,p}) and (\textit{n,d}) cross sections still present the unresolved inconsistencies and cover a limited energy range. The (\textit{n,p}) and (\textit{n,d}) reactions on natural carbon consist of the following relevant components: $^{12}$C(\textit{n,p})$^{12}$B, $^{13}$C(\textit{n,p})$^{13}$B, $^{12}$C(\textit{n,d})$^{11}$B and $^{13}$C(\textit{n,d})$^{12}$B. Apart from the emission of proton or deuteron, the short-lived product nuclei $^{12}$B and $^{13}$B, with half-lives of 20.2~ms and 17.3~ms against $\beta$ decay, respectively, are followed by the emission of high energy electrons with average energy around 6~MeV. The (\textit{n},cp) reactions are thus also important in radioprotection calculations in accelerator facilities where a field of high-energy neutrons is present. They include accelerator-based neutron facilities, spallation neutron sources, or neutron irradiation facilities for fusion related material research, such as the IFMIF-DONES facility now under construction. Finally, cross sections for (\textit{n,p}) and (\textit{n,d}) reactions on carbon play a role in dosimetry for space applications, while the increasing importance of diamond devices as neutron detectors -- in particular as neutron-flux monitors -- in the fast energy region calls for accurate data on these reactions in a wide energy range.

Apart from being important for several applications, new data on neutron-induced reactions on carbon may help validate and refine theoretical models. For example, this is the case with calculations performed with the TALYS code~\cite{talys_2012,talys_2023,talys_web}, that relies on different formalisms and parameter choices for predictions of reaction cross sections. Given the energy range of interest, for the neutron-induced reactions on carbon above 14~MeV, the relevant models included in TALYS are related to the direct and pre-equilibrium mechanisms. It is therefore justified to use such code since the main (or all) important reactions are included in various nuclear model representations (e.g. exciton model, coupled channel formalism) and databases (e.g. level density). The verification of used assumptions, and the eventual optimization of the code can only be achieved by comparing calculations with experimental data.

Despite the importance of the $^\car$C(\textit{n,p}) and $^\car$C(\textit{n,d}) reactions for fundamental and applied nuclear physics, only few datasets are available in literature up to date, exhibiting large discrepancies between each other, and covering a limited energy range of just a few MeV above the reaction threshold. The main reason for the lack of reliable data is related to the neutron energy range of most existing facilities, mostly limited to just above 20~MeV, only a few MeV above the reaction threshold of 13.6~MeV and 14.9~MeV for the (\textit{n,p}) and (\textit{n,d}) reaction respectively. Aside from the scarcity and inconsistency of the experimental data, the evaluated data from major libraries seem to be inconsistent with the theoretical predictions from codes such as TALYS.

Important information on the integral cross section of the $^{12}$C(\textit{n,p})$^{12}$B reaction was obtained a few years ago at n\_TOF, by means of in-beam activation analysis~\cite{carbon_prc,carbon_epja}, i.e. by counting the number of the produced $^{12}$B nuclei. The results indicated that most major libraries grossly underestimated the cross section of this reaction. The measurement, however, did not provide direct indication on the energy dependence of the cross section.

In order to further investigate this reaction, a new measurement was proposed at n\_TOF, aiming at the determination of the energy dependence of the $^\car$C(\textit{n,p}) and $^\car$C(\textit{n,d}) cross sections in a wider energy range than currently available. To this end a new detection system was developed, based on $\Delta E$-$E$ telescopes made of position sensitive silicon detectors. The results of the measurement are reported here. In Section~\ref{setup} the experimental setup is described together with the main steps of the data analysis. The results and comparisons with previous data and current evaluations are presented in Section~\ref{results}, together with a discussion on the implications of the new results on theoretical calculations. Conclusions are finally given in Section~\ref{conclusions}.

\section{Experimental setup and data analysis }
\label{setup}

The measurement was performed in the first experimental area (EAR1) at n\_TOF, CERN, located at 185~m distance from the spallation target. The details on the facility and on the neutron beam in EAR1 can be found in Refs.~\cite{ntof,flux_ear1}. Specifically, the neutron flux above 10~MeV is determined by the dedicated measurements using Parallel Plate Avalanche Counters relying on the $^{235}$U(\textit{n,f}) reaction~\cite{flux_ear1}. Within the energy range of interest (15~MeV -- 25~MeV) the flux varies between approximately 800--1300 neutrons per MeV per nominal neutron pulse, and its detailed parameterizaton was used during the subsequent data analysis.

\begin{figure}[t!]
\centering 
\includegraphics[width=0.45\linewidth]{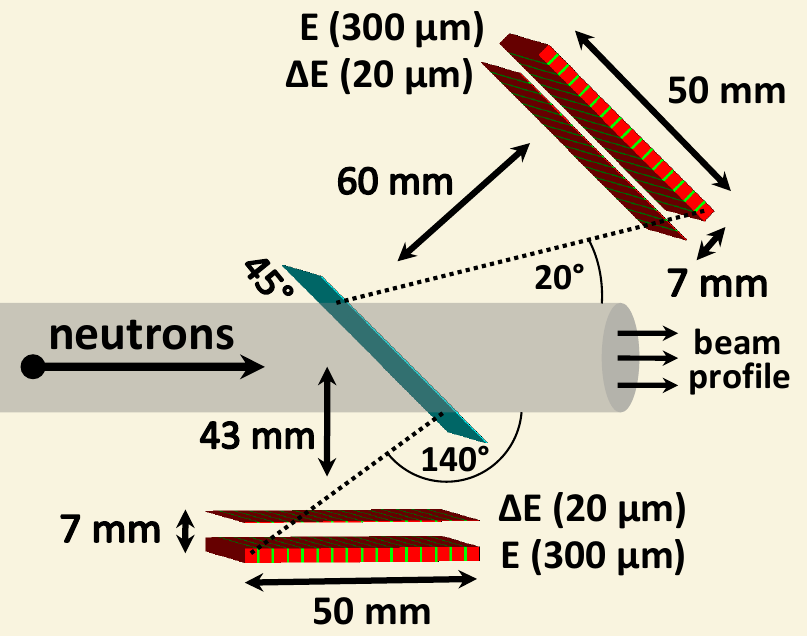}	
\caption{A schematic drawing of the experimental setup, with the two telescopes made of position-sensitive silicon detectors.} 
\label{fig_setup}%
\end{figure}

 The detection setup consisted of two $\Delta E$-$E$ telescopes made of two position-sensitive silicon detectors: a 20~$\mu$m thick one, acting as $\Delta E$ detector, and a 300~$\mu$m thick one for $E$~reconstruction. Each detector was $5\,\mathrm{cm}\times5\,\mathrm{cm}$ in area, and segmented in 16~strips, each 3~mm wide. The two detectors were mounted with the strips oriented in the same direction and perpendicular to the beam. A schematic drawing of the setup is shown in Figure~\ref{fig_setup}. The two telescopes covered the angular range between 20$^\circ$ and 140$^\circ$.  More details on experimental setup can be found in Ref.~\cite{site}.

The detectors were read-out by custom made preamplifiers, and the signals acquired by the n\_TOF data acquisition system based on flash ADC. A sampling rate of 125~MS/s was chosen for these detectors, in a 14-bit resolution.
A $^\car$C sample of rectangular shape, with $5\,\mathrm{cm}\times5\,\mathrm{cm}$ area and 0.25~mm thickness, was mounted in the vacuum chamber, at 45$^\circ$ angle relative to the beam direction, as shown in Fig.~\ref{fig_setup}, in order to optimize between the effective sample thickness crossed by the neutron beam and the energy loss of the reaction products in the sample, emitted towards the two telescopes. Within the energy range of interest most of the neutron beam profile is contained within a radius of 1.2~mm from the beam axis. The carbon sample covered the entire beam and no special correction had to be performed. The acquired detector signals were analysed by means of the standard n\_TOF Pulse Shape reconstruction procedure, based on pulse shape fitting~\cite{psa}. Extracted information included the signal timing and amplitude. A method for optimal synchronization of signals from separate silicon strips was developed in Ref.~\cite{sync}.

For high-performance particle identification, individual $\Delta E$-$E$ pairs of silicon strips were analysed, separately, in the 3D parameter space ($E_n, \Delta E, E$), with the neutron energy $E_n$ as an additional parameter alongside the deposited energies~$\Delta E$ and~$E$. To this end, a reconstruction procedure based on neural network was employed. The procedure is described in detail in Ref.~\cite{neural}. Monte Carlo simulation of the energy loss by different particles in the $\Delta E$-$E$ telescope, folded by the measured detector resolution, was used to train the neural network, resulting in a high degree of discrimination between protons (including punch-through ones), deuterons and tritons, all the way up to neutron energy of 25~MeV. Above this energy, the patterns related to the different particles start to mix among each other, introducing a degree of uncertainty in the particle identification.

One of the main problems in the measurement is associated with the angular distribution of the emitted reaction products, which is not expected to be isotropic. A further complication is that such a distribution depends on the excited state of the residual nucleus populated in the reaction. Therefore, the determination of the (\textit{n,p}) or (\textit{n,d}) cross sections at a given neutron energy depends on the branching ratios for populating the accessible excited states of the residual nucleus, as well as the corresponding angular distributions of the reaction products. In addition, since a $^\car$C sample was used in the measurement, the analysis procedure has to take into account the isotopic composition of the sample. Although the natural abundance of the $^{13}$C isotope is small (only~1.1\%), in the efficiency corrections the contribution of the (\textit{n,p}) and (\textit{n,d}) reactions from $^{13}$C has to be included, since it is not possible to discriminate the contribution of different carbon isotopes to the measured cross section.

The reaction cross section can be extracted by appropriately linearising and inverting the following expression~\cite{angular}: 
\begin{equation}
\frac{\D N_{ij}}{\D E_\mathrm{tof}}=\sum_{C} \eta_C \int_0^\infty \!\!\! R(E_\mathrm{tof},E_n) \phi(E_n) F_{i,j;C}(E_n) \sigma_C(E_n) \D E_n .
\label{master_eq}
\end{equation}
Here, $i$ and $j$ refer to the particular pair of $\Delta E$-$E$ strips, $N_{ij}$ to the number of counts in that pair, $E_\mathrm{tof}$ to the neutron energy reconstructed from the measured time-of-flight, with $E_n$ as the true neutron energy. $R(E_\mathrm{tof},E_n)$ is the resolution function of the n\_TOF neutron beam, $\phi(E_n)$ the neutron flux (for both see Refs.~\cite{ntof,flux_ear1,beam}), $C\in\{{^{12}\mathrm{C}},{^{13}\mathrm{C}}\}$ the carbon isotope, $\eta_C$ the corresponding areal density (in atoms/barn) and $\sigma_C(E_n)$ the reaction cross section for a given isotope, at a given neutron energy. Neutron flux is one of the two main sources of the systematic uncertainties. Within the energy range of interest the systematic uncertainty in the evaluated n\_TOF flux amounts to 3\%~\cite{flux_ear1}, affecting in equal measure the uncertainties in the extracted cross sections. The correction factors $F_{i,j;C}$~\cite{angular}:
\begin{equation}
\begin{split}
F_{ij;C}(E_n)= \sum_x \rho_C(x,E_n)\int_{-1}^1 \varepsilon_{ij;C}(x,E_n,\cos\theta) A_C(x,E_n,\cos\theta)\,\D(\cos\theta)
 \end{split}
\end{equation}
are related to the efficiency of the experimental setup. The central quantities affecting them are the efficiencies $\varepsilon_{ij;C}(\mathrm{x},E_n,\cos\theta)$ for detecting the protons or deuterons emitted under an angle $\theta$ relative to a neutron beam direction, leaving a residual boron nucleus (determined by a specific carbon isotope~$C$) in any of the energetically accessible states, indexed by~$x$. These efficiencies represent the probability for the protons or deuterons to be detected, following their emission inside the sample. Thus, they account for any loss of counts due to their passage through matter. Due to the angular sensitivity, one needs to take into account the angular distributions $A_C(x,E_n,\cos\theta)$ of the reaction products. A consideration of the branching ratios $\rho_C(x,E_n)$ is also required due to the sensitivity to a particular state of a residual nucleus.

The detection efficiencies $\varepsilon_{i,j;C}$ were determined by means of Monte Carlo simulations performed with the GEANT4 toolkit~\cite{geant4_1,geant4_2,geant4_3}, in which a detailed software replica of the experimental setup has been implemented. In the simulations, protons and deuterons are generated with a given energy and angular distribution, and propagated inside the apparatus. The simulated energy and position, suitably folded by the resolution of the setup, are then used to determine the detection efficiency. The correction factors $F_{i,j;C}$ can be obtained as a function of neutron energy, by generating the emitted particles according to the energy and angular distribution corresponding to the different excited states of the residual nucleus populated at that neutron energy. However, neither the data on the branching ratios for different states of the residual nucleus, nor the data on the angular distribution of emitted particles are available from past experiments. In principle, some information about the angular distribution of the reaction products could be extracted from the n\_TOF data. However, our statistics is not sufficient for obtaining this angular information with a reasonable degree of statistical confidence. As a consequence, one needs to rely on the theoretical estimates of the branching ratios and angular distributions. A key part of the work is therefore devoted to estimating the uncertainty in the correction factors associated with the assumed branching ratios and angular distributions.

In this work the TALYS-2.0 code~\cite{talys_2012,talys_2023,talys_web} was used to generate the branching ratios for the individual states of residual nuclei, and angular distributions of the emitted protons and deuterons. These theoretical distributions were then used in combination with GEANT4 simulations to determine the efficiency correction factors.

The adoption of particular TALYS models is the second major source of the systematic uncertainties. In order to assess these model uncertainties related to the theoretical calculations, different models and assumptions were used. Various models can be selected in TALYS for different quantities with a range of possibilities from a number of phenomenological and (semi-)microscopic models. A list of model parameters used in TALYS is reported in Table~\ref{tab_models}. In accordance with Refs.~\cite{tendl_astro_web,tendl_astro}, we refer to each particular combination of model parameters as a model set. The meaning of parameters may be found in Refs.~\cite{tendl_astro,talys_keff}. For the present work, 384 model sets were used in TALYS-2.0 calculations, generating a corresponding number of angular distributions and branching ratios. Each TALYS model set separately provided branching ratios and angular distributions of the emitted products from neutron-induced reactions on both $^{12}$C and $^{13}$C.

\begin{table}[b!]
\caption{List of TALYS model parameters and their values used for generating 384 model sets (each model set is a specific combination of model parameters). Not all combinations of parameter values were used, as indicated by their shared superscript letters (a, b). Thus, only the combinations (8,8) and (9,3) were used for (strength,strengthm1) pair of parameters. Each value without a superscript was combined with all values of other parameters. A description of parameters may be found in Refs.~\cite{tendl_astro,talys_keff}.} 
\centering

\begin{tabular}{cccccc}
\hline\hline
Parameter & Values & Parameter & Values & Parameter & Values\\
\hline
strength & 8$^\mathrm{a}$, 9$^\mathrm{b}$ & jlmomp & n, y & massmodel & 0, 1, 2, 3\\
strengthm1 & 3$^\mathrm{b}$, 8$^\mathrm{a}$  & colenhance & n, y & alphaomp & 5, 6\\
ldmodel & 1, 2 & widthmode & 0, 1, 2 & fismodel & 1 \\
\hline\hline
\end{tabular}

\label{tab_models}
\end{table}

Finally, Eq.~(\ref{master_eq}) was used to extract the cross sections for the (\textit{n,p}) and (\textit{n,d}) reactions on $^\car$C from the measured count rates, relying on a particular model set for the efficiency correction. An unweighted, arithmetic mean of the cross sections resulting from all 384 different model sets finally provided the average cross section.

In order to estimate the model uncertainties in the extracted cross sections, related to the angular distributions and branching ratios, a further analysis was performed. We considered two additional extreme assumptions: isotropic angular distributions (for all excited states and at all energies) and artificially constructed branching ratios. Analysis was then repeated by combining the isotropic angular distributions with all aforementioned 384 branching ratio dependences from TALYS-2.0; by combining the artificial branching ratios with 384 angular distribution sets from TALYS-2.0; and by combining the isotropic angular distributions with artificial branching ratios (a single fully artificial combination). The procedure briefly outlined above will be described in detail in a forthcoming paper. 

\begin{figure}[t!]
\centering 
\includegraphics[width=0.5\linewidth]{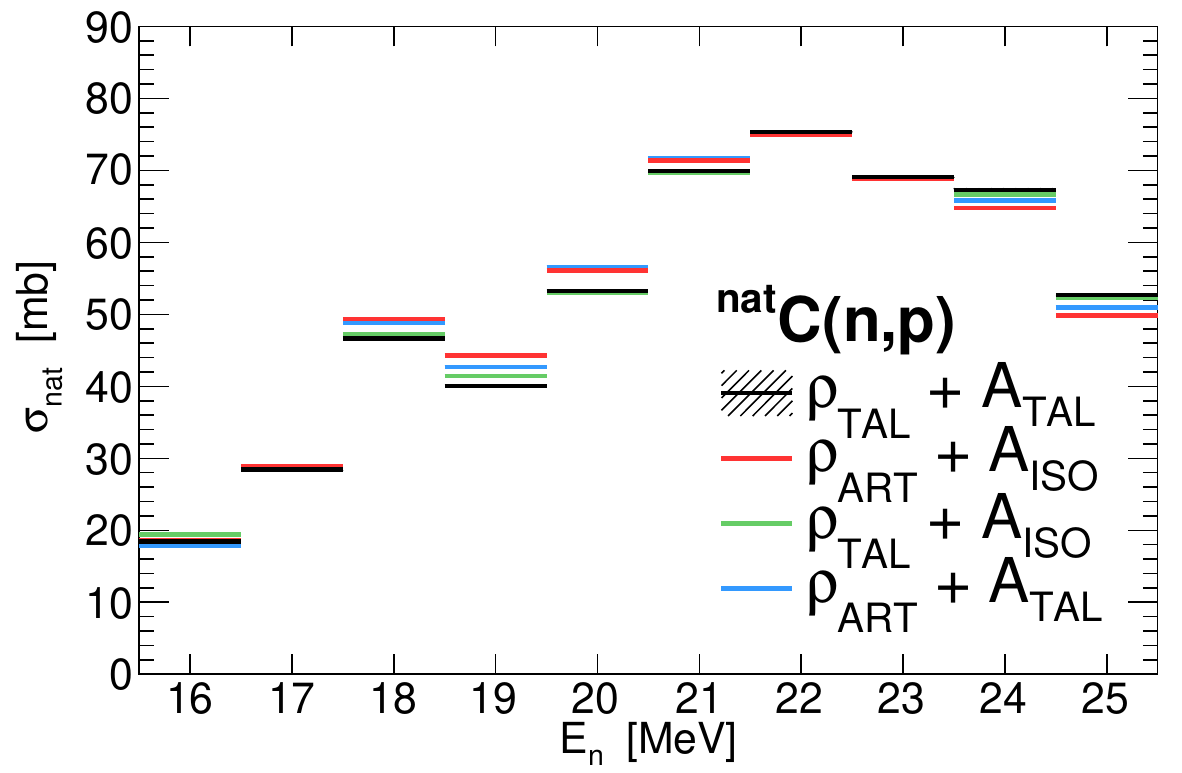}\includegraphics[width=0.5\linewidth]{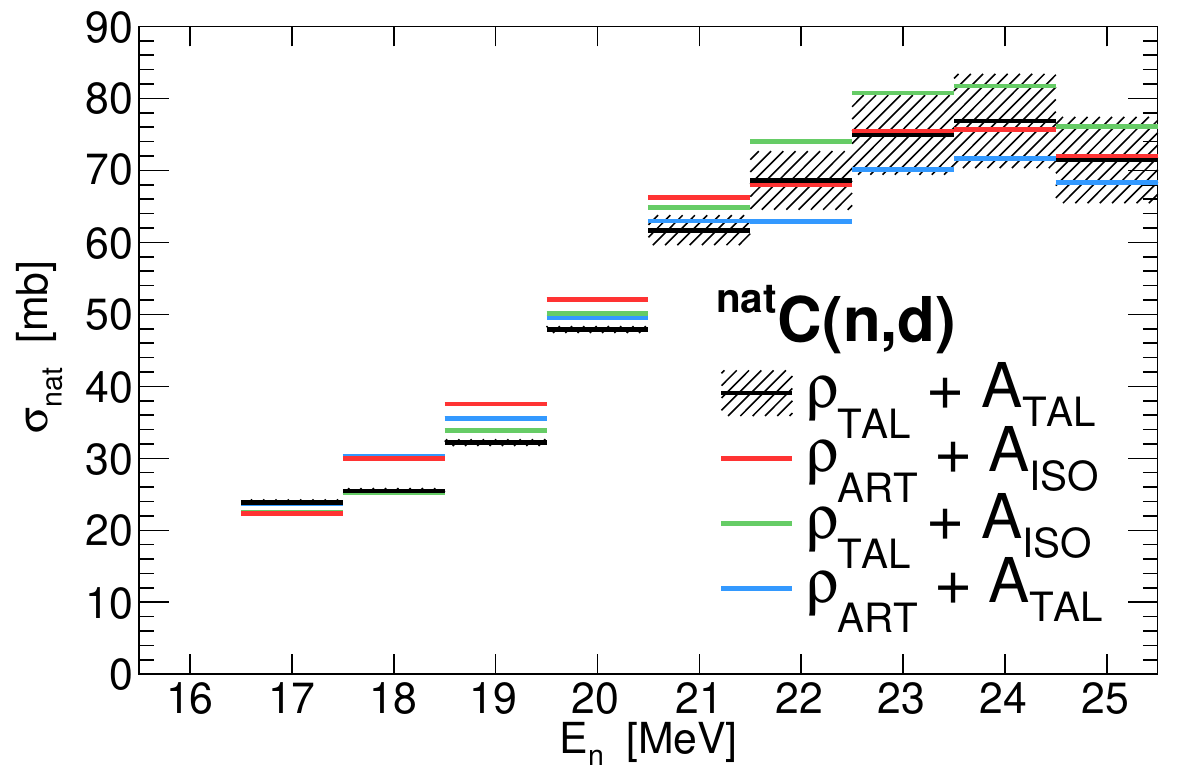}
\caption{$^\car$C(\textit{n,p}) and $^\car$C(\textit{n,d}) cross sections reconstructed using different combinations of angular distributions and branching ratios, including those from TALYS-2.0 ($A_\mathrm{TAL}$ and $\rho_\mathrm{TAL}$), the isotropic angular distributions ($A_\mathrm{ISO}$) and the artificial branching ratios ($\rho_\mathrm{ART}$). The shaded band shows the Root Mean Square of the results from 384 TALYS model sets.}
\label{fig_cross_sec}%
\end{figure}

The resulting cross sections, reconstructed by using the described combinations of angular distributions~$A$ and branching ratios~$\rho$, are shown in Figure~\ref{fig_cross_sec}. Horizontal lines show the averages over all combinations of the same type (i.e. over 384 $\rho_\mathrm{TAL}+A_\mathrm{TAL}$ combinations; over 384 $\rho_\mathrm{ART}+A_\mathrm{TAL}$ and 384 $\rho_\mathrm{TAL}+A_\mathrm{ISO}$ combinations; and a single $\rho_\mathrm{ART}+A_\mathrm{ISO}$ combination). The shaded bands in the figure represent only the Root Mean Square of the cross sections reconstructed using 384 model sets from TALYS-2.0. It should be noted that for the (\textit{n,p}) reaction those uncertainties are so small (well below 1\%) that they are not visible. A larger uncertainty due to a spread of TALYS-2.0 results -- of up to 9\% -- characterizes the extracted cross section for the (\textit{n,d}) reaction. This is due to a larger sensitivity of the TALYS calculations to the model parameters for this particular reaction.

The results obtained with isotropic angular distributions and artificial branching ratios were used only in the estimation of the model uncertainties. Their deviation from the finally adopted cross section values -- obtained by averaging 384 results obtained from TALYS model sets -- was considered alongside the Root Mean Square of 384 TALYS model sets. A conservative model uncertainty estimate was obtained by taking a maximum value between the RMS of 384 results obtained with TALYS ($\Delta \sigma_\mathrm{talys}$), and the largest deviation $\delta \sigma_\mathrm{max}$ of the averages indicated in Fig.~\ref{fig_cross_sec} by colored lines from the adopted cross section averages (black lines): \mbox{$\Delta \sigma_\mathrm{model}=\max(\Delta \sigma_\mathrm{talys},\delta \sigma_\mathrm{max})$}.

\section{Results and discussion}
\label{results}

Reaction thresholds for both the $^{12}$C(\textit{n,p}) and $^{13}$C(\textit{n,p}) reaction are approximately 13.6~MeV. Reaction thresholds for the $^{12}$C(\textit{n,d}) and $^{13}$C(\textit{n,d}) reactions are 14.9~MeV and 16.5~MeV, respectively. The thresholds for the natural carbon are determined by the lowest threshold between $^{12}$C and $^{13}$C: 13.6~MeV for the $^\car$C(\textit{n,p}) and 14.9~MeV for the $^\car$C(\textit{n,d}) reaction. The detection thresholds of $\Delta E$-$E$ telescopes for both types of reactions are approximately 1~MeV above their reaction thresholds. Furthermore, just above the detection thresholds -- up to about 15.5~MeV for (\textit{n,p}) and 16.5~MeV for (\textit{n,d}) -- the data suffer from a very low detection statistics. For this reason we report in this work the (\textit{n,p}) data starting from the energy bin centered at 16~MeV and the (\textit{n,d}) data from a bin centered at 17~MeV.

Table~\ref{tab_results} lists the final results for the $^\car$C(\textit{n,p}) and $^\car$C(\textit{n,d}) cross sections. The model~($\Delta \sigma_\mathrm{model}$) and flux~($\Delta \sigma_\mathrm{flux}$) contributions to the systematic uncertainty~($\Delta \sigma_\mathrm{sys}^2=\Delta \sigma_\mathrm{model}^2+\Delta \sigma_\mathrm{flux}^2$) are clearly separated. Systematic and statistical uncertainties~($\Delta \sigma_\mathrm{stat}$) yield the total uncertainties ($\Delta \sigma_\mathrm{tot}^2=\Delta \sigma_\mathrm{stat}^2+\Delta \sigma_\mathrm{sys}^2$).

\begin{table*}[t!]
\caption{Energy dependent cross section for the $^\car$C(\textit{n,p}) and $^\car$C(\textit{n,d}) reactions from n\_TOF. Listed uncertainties are: statistical~($\Delta \sigma_\mathrm{stat}$), model related~($\Delta \sigma_\mathrm{model}$), flux related ($\Delta \sigma_\mathrm{flux}$; fixed at 3\%), systematic ($\Delta \sigma_\mathrm{sys}^2=\Delta \sigma_\mathrm{model}^2+\Delta \sigma_\mathrm{flux}^2$) and total ($\Delta \sigma_\mathrm{tot}^2=\Delta \sigma_\mathrm{stat}^2+\Delta \sigma_\mathrm{sys}^2$). All are given in absolute and in relative values.}
\centering

\begin{tabular}{ccccccc}
\hline\hline
\multicolumn{7}{c}{\vrt$^{\car}$C(\textit{n,p})\vrt} \\
\hline
$E_n$ (MeV) & $\sigma_\car$ (mb) & $\Delta \sigma_\mathrm{stat}$ (mb) & $\Delta \sigma_\mathrm{model}$ (mb) & $\Delta \sigma_\mathrm{flux}$ (mb) & $\Delta \sigma_\mathrm{sys}$ (mb) & $\Delta \sigma_\mathrm{tot}$ (mb)\\
\hline
16 & {18.4} & {1.8} (9.8\%) & {1.0} (5.4\%) & 0.6 (3.0\%) & 1.2 (6.5\%) & 2.2 (12.0\%) \\
17 & {28.4} & {1.5} (5.3\%) & {0.5} (1.8\%) & 0.9 (3.0\%) & 1.0 (3.5\%) & 1.8 (6.3\%) \\
18 & {46.6} & {1.6} (3.4\%) & {2.7} (5.8\%) & 1.4 (3.0\%) & 3.1 (6.7\%) & 3.4 (7.3\%) \\
19 & {40.0} & {1.3} (3.3\%) & {4.2} (10.5\%) & 1.2 (3.0\%) & 4.4 (11.0\%) & 4.6 (11.5\%) \\
20 & {53.3} & {1.5} (2.8\%) & {3.2} (6.0\%) & 1.6 (3.0\%) & 3.6 (6.8\%) & 3.9 (7.3\%) \\
21 & {69.9} & {1.6} (2.3\%) & {1.8} (2.6\%) & 2.1 (3.0\%) & 2.8 (4.0\%) & 3.2 (4.6\%) \\
22 & {75.3} & {1.6} (2.1\%) & {0.3} (0.4\%) & 2.3 (3.0\%) & 2.3 (3.1\%) & 2.8 (3.7\%) \\
23 & {69.1} & {1.5} (2.2\%) & {0.3} (0.4\%) & 2.1 (3.0\%) & 2.1 (3.0\%) & 2.6 (3.8\%) \\
24 & {67.3} & {1.4} (2.1\%) & {2.5} (3.7\%) & 2.0 (3.0\%) & 3.2 (4.8\%) & 3.5 (5.2\%) \\
25 & {52.7} & {1.3} (2.5\%) & {2.8} (5.3\%) & 1.6 (3.0\%) & 3.3 (6.3\%) & 3.5 (6.6\%) \\
\hline\hline\\
\hline\hline
\multicolumn{7}{c}{\vrt$^{\car}$C(\textit{n,d})\vrt} \\
\hline
$E_n$ (MeV) & $\sigma_\car$ (mb) & $\Delta \sigma_\mathrm{stat}$ (mb) & $\Delta \sigma_\mathrm{model}$ (mb) & $\Delta \sigma_\mathrm{flux}$ (mb) & $\Delta \sigma_\mathrm{sys}$ (mb) & $\Delta \sigma_\mathrm{tot}$ (mb)\\
\hline
17 & {23.9} & {2.3} (9.6\%) & {1.6} (6.7\%) & 0.7 (3.0\%) & 1.7 (7.1\%) & 2.8 (11.7\%) \\
18 & {25.4} & {1.4} (5.5\%) & {4.8} (18.9\%) & 0.8 (3.0\%) & 4.9 (19.3\%) & 5.1 (20.1\%) \\
19 & {32.2} & {1.3} (4.0\%) & {5.4} (16.8\%) & 1.0 (3.0\%) & 5.5 (17.1\%) & 5.6 (17.4\%) \\
20 & {47.9} & {1.4} (2.9\%) & {4.1} (8.6\%) & 1.4 (3.0\%) & 4.4 (9.2\%) & 4.6 (9.6\%) \\
21 & {61.7} & {1.6} (2.6\%) & {4.6} (7.5\%) & 1.9 (3.0\%) & 4.9 (7.9\%) & 5.2 (8.4\%) \\
22 & {68.6} & {1.7} (2.5\%) & {5.7} (8.3\%) & 2.1 (3.0\%) & 6.0 (8.7\%) & 6.2 (9.0\%) \\
23 & {74.9} & {1.7} (2.3\%) & {5.8} (7.7\%) & 2.2 (3.0\%) & 6.2 (8.3\%) & 6.5 (8.7\%) \\
24 & {76.9} & {1.7} (2.2\%) & {6.5} (8.5\%) & 2.3 (3.0\%) & 6.9 (9.0\%) & 7.2 (9.4\%) \\
25 & {71.4} & {1.7} (2.4\%) & {6.0} (8.4\%) & 2.1 (3.0\%) & 6.4 (9.0\%) & 6.6 (9.2\%) \\
\hline\hline
\end{tabular}

\label{tab_results}
\end{table*}

\begin{figure}[b!]
\centering 
\includegraphics[width=0.5\linewidth]{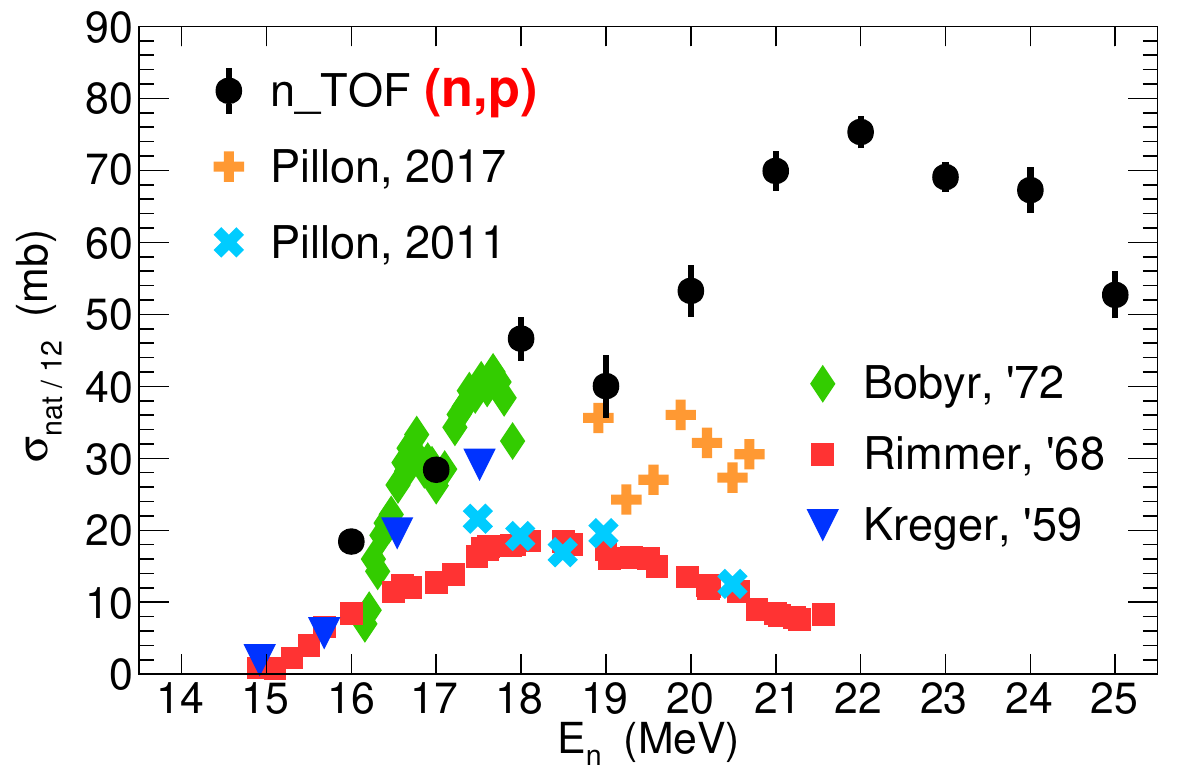}\includegraphics[width=0.5\linewidth]{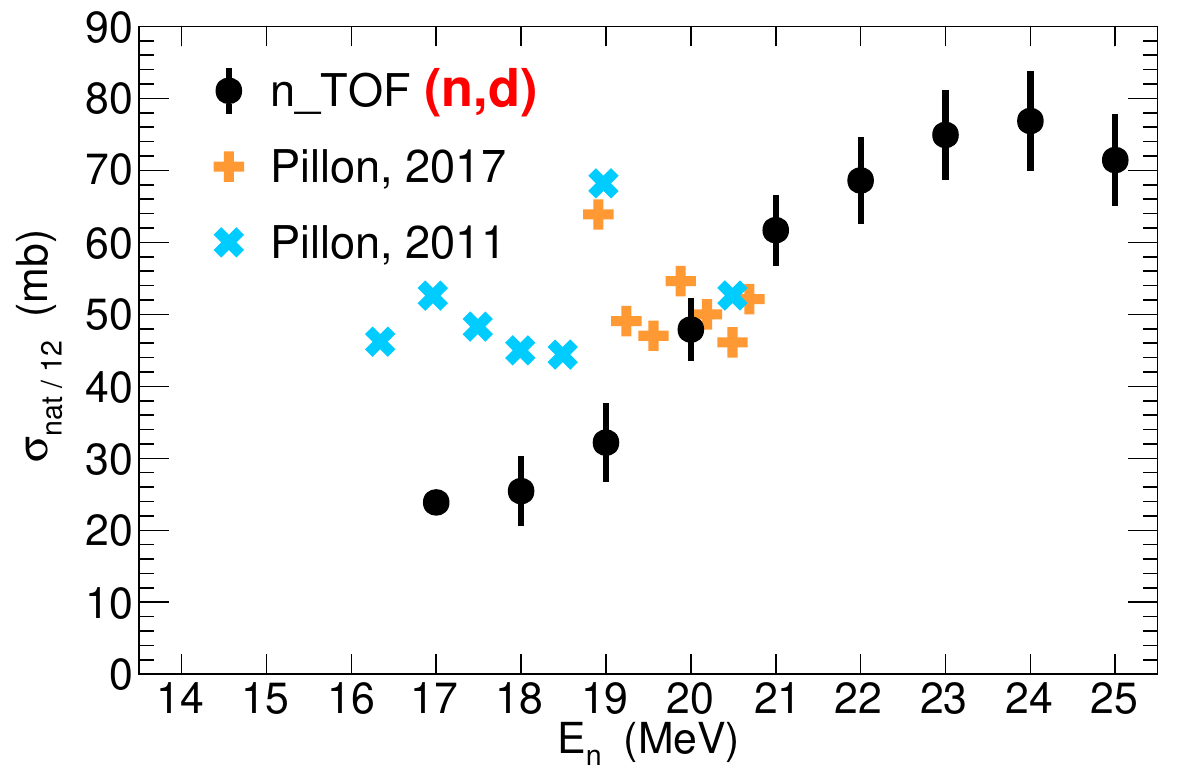}
\caption{n\_TOF cross sections for the $^\car$C(\textit{n,p}) and $^\car$C(\textit{n,d}) reactions compared to the previous experimental data. Systematic uncertainties are shown for n\_TOF data. The subscript from $\sigma_{\car/12}$ means that some of the experimental data either explicitly refer to $^{12}$C or are listed under $^{12}$C in the available databases.} 
\label{fig_exfor}%
\end{figure}

The final n\_TOF results are compared to the previous experimental data in Figure ~\ref{fig_exfor}. The error bars represent the systematic uncertainties only (the statistical uncertainties are in most cases smaller than the point size). The present $^\car$C(\textit{n,p}) data agree only with the datasets of Kreger and Kern~\cite{kreger_1959} and Bobyr et al.~\cite{bobyr_1972}, while a large difference is observed relative to the results of Rimmer and Fisher~\cite{rimmer_1968} and both data sets of Pillon et al.~\cite{pillon_2011,pillon_2017}. Their 2011 dataset is only sensitive to the ground state, and second and third excited state in $^{12}$B residual nucleus. A second dataset, from 2017, provides higher cross sections, but with a flat behavior, incompatible with the trend observed in the n\_TOF measurement. Although they only report the partial cross sections for a few lowest states in the residual nucleus, within first $\approx$4~MeV the total cross section is fully determined by these states. The most recent measurements of the (\textit{n,p}) reaction include those by Majerle et al.~\cite{majerle} and Kuvin et al.~\cite{kuvin}. Essentially, they provide the partial cross sections for the reaction channel populating only the ground state of the residual nucleus. As such, they are not directly comparable to the n\_TOF results. 
A behavior similar to the $^\car$C(\textit{n,p}) reaction is observed for the $^\car$C(\textit{n,d}) one. In this case there are only two previous datasets, both from Pillon et al.~\cite{pillon_2011,pillon_2017}, which are in good agreement with the present n\_TOF data only around 20~MeV, at the upper limit of those measurements. 

\pagebreak

\begin{figure}[t!]
\centering 
\includegraphics[width=0.5\linewidth]{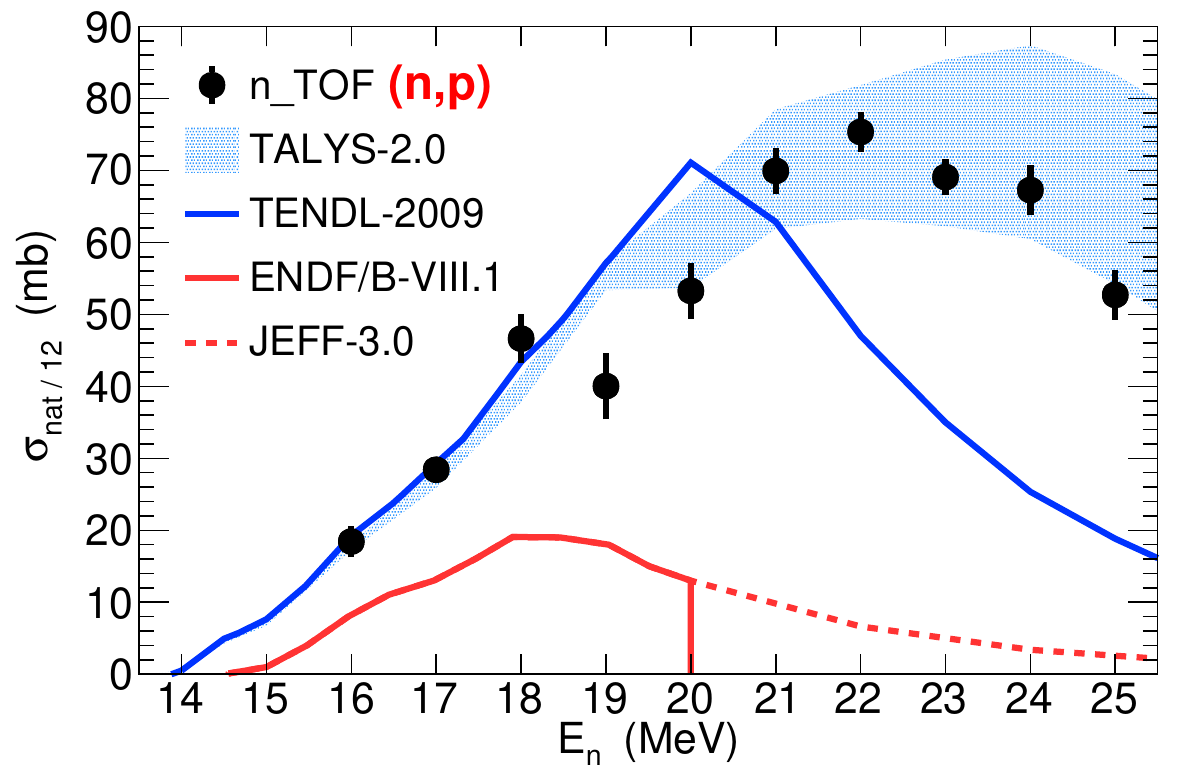}\includegraphics[width=0.5\linewidth]{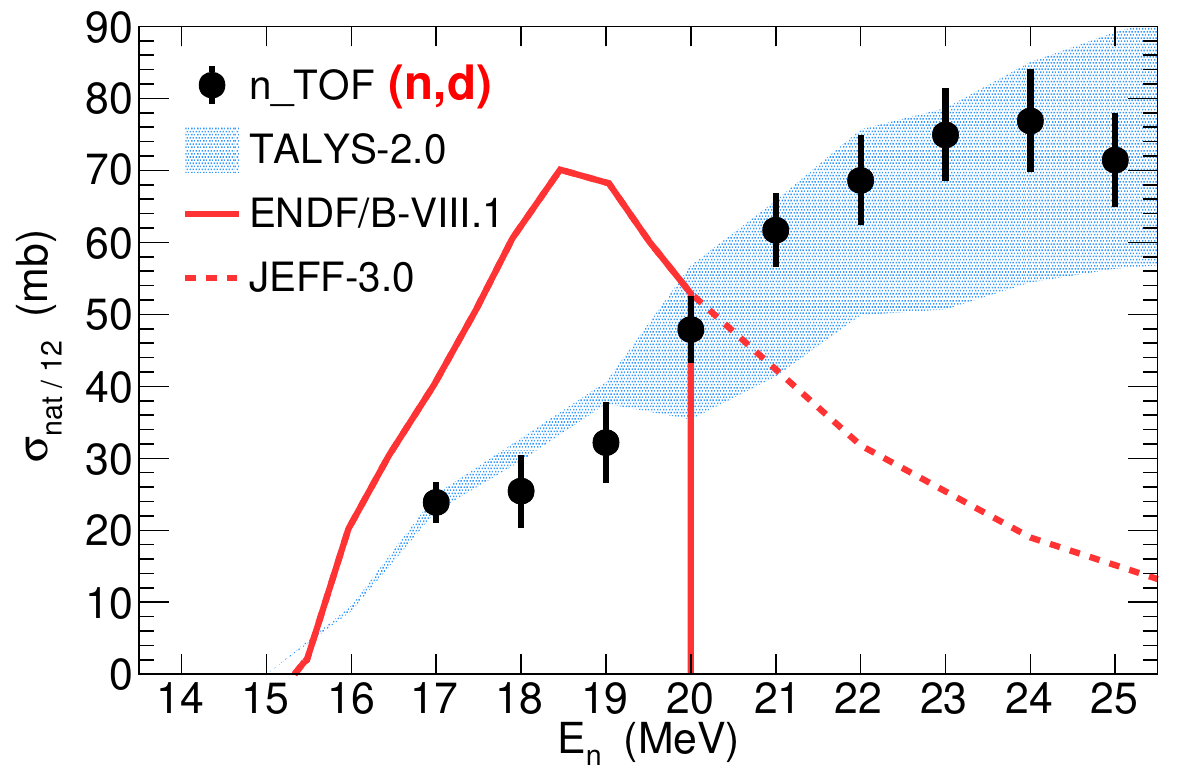}
\caption{n\_TOF cross sections for the $^\car$C(\textit{n,p}) and $^\car$C(\textit{n,d}) reactions compared to the relevant evaluation libraries and to the range of TALYS-2.0 calculations. ENDF/B-VIII.1 is representative of most of the other major libraries. Total uncertainties are shown for the n\_TOF data (see Tab.~\ref{tab_results}).} 
\label{fig_endf}%
\end{figure}

Figure~\ref{fig_endf} compares the present data with the cross sections from the relevant evaluation libraries and TALYS-2.0 calculations. For TALYS predictions, the band spans a range between the minimum and maximum cross sections calculated with all 384 model sets previously discussed. For the $^\car$C(\textit{n,p}) reaction a large discrepancy is observed, both in shape and absolute value, with the evaluated cross sections in ENDF-B/VIII.1. We remind here that up to 20~MeV, this cross section (provided by vast majority of evaluation libraries) is purely based on the dataset from Rimmer and Fisher~\cite{rimmer_1968}, thus explaining the observed difference. The only exception is TENDL-2009 which is based on TALYS-1.2 calculations~\cite{tendl_2009}. For this library the cross section below 20~MeV is a factor of 3 higher relative to all other evaluations. It should be noted, however, that some of the later versions of TENDL (TENDL-2021 and TENDL-2023) have adopted the ENDF evaluation, and as such now grossly underestimate the cross section measured at n\_TOF. Another problem of the evaluated cross sections is that the energy range from most of the libraries is limited to 20~MeV, thus missing the peak of the cross section, as indicated by the present results. JEFF-3.0 and BROND-2.2 provide an extension of the ENDF evaluation up to 32~MeV, which is in drastic disagreement with the n\_TOF data (the later versions of these databases, including the latest JEFF-3.3 and BROND-3.1 stop at 20~MeV). Regarding the $^\car$C(\textit{n,d}) reaction, the cross sections in ENDF/B-VIII.1 (and in all other major libraries) are clearly incompatible with the present results.

Contrary to all aforementioned major libraries, a good agreement is observed between the n\_TOF results and the TALYS-2.0 calculations, although the shape is different, with the peak in TALYS calculations reached around 24~MeV for the (\textit{n,p}) and 29~MeV for the (\textit{n,d}) reaction. It should be mentioned that by no means can this agreement be attributed to the use of TALYS calculations for angular distribution and branching ratio corrections. Indeed, at no point were the absolute values of the TALYS cross section used in the analysis of the experimental data. After all, similar results are obtained even when using fully artificial distributions (isotropic angular distributions and artificial branching ratios; see Fig.~\ref{fig_cross_sec}). As a consequence, the good agreement between n\_TOF results and TALYS-2.0 theoretical calculations represents a strong indication of the predictive power of the code for these reactions in the tested energy range.

\section{Conclusions}
\label{conclusions}

The cross section for the $^\car$C(\textit{n,p}) and $^\car$C(\textit{n,d}) reactions has been measured at n\_TOF in a wide neutron energy range, covering for the first time the region up to 25~MeV. The experimental setup for light charged particle detection and identification consisted of two silicon-strip $\Delta E$-$E$ telescopes, mounted inside a vacuum chamber and located in the first experimental area of the n\_TOF facility. In order to extract the cross section from the number of protons and deuterons detected in the telescope, a correction factor related to the detection efficiency had to be determined. As the efficiency essentially depends on the energy and angular distribution of the emitted particles, which in turn depend on the excited state of the residual nucleus populated in the reaction, assumptions had to be made on the branching ratios for the energetically available levels of the involved boron nuclei. To this end, extensive theoretical calculations performed with the code TALYS-2.0 were used to generate 384 sets of energy dependent branching ratios and corresponding angular distributions, by combining different physics models available in the code.

For both reactions the present data point to a maximum value of the cross section around 76~mb, peaking around 22~MeV and 24~MeV for the (\textit{n,p}) and (\textit{n,d}) reaction, respectively. In case of the $^\car$C(\textit{n,p}) reaction the maximum cross section is significantly larger than suggested by the major evaluation libraries, fully consistent with a previous finding based on an integral measurement at n\_TOF. A disagreement between the present data and all major libraries -- regarding both the shape and a magnitude of the cross section -- calls for a complete reevaluation of the (\textit{n,p}) and (\textit{n,d}) cross sections for $^\car$C.

Finally, a good agreement is observed between the present data and model calculations performed with TALYS-2.0. In fact, the n\_TOF cross section falls within the range of predictions that can be obtained with different choices of models and model parameters. Nevertheless, for both reactions the shape of the predicted cross section seems to be somewhat different, with a maximum cross section in TALYS-2.0 reached at higher neutron energies -- around 24~MeV and 29~MeV for the (\textit{n,p}) and (\textit{n,d}) reaction, respectively. A reasonable agreement between the TALYS predictions and the measured data provides a strong indication that the code could suitably be used even for light nuclei, in a given neutron energy range. Furthermore, the present data could be useful to refine those calculations, by fine-tuning the models and related parameters.

\section*{Acknowledgements}

This work was supported by the Croatian Science Foundation under the project number HRZZ-IP-2022-10-3878, and by the National Science Centre in Poland (Grant No. UMO-2021/41/B/ST2/00326). This project has received funding from the European Union’s Horizon Europe Research and Innovation programme under Grant Agreement No 101057511. D.~Rochman acknowledges support from EU APRENDE project, number 101164596. Geant4 simulations were run at the Laboratory for Advanced Computing, Faculty of Science, University of Zagreb. Support from the funding agencies of all participating institutes are also gratefully acknowledged.








\bibliographystyle{elsarticle-harv} 


\end{document}